\documentclass[twocolumn,superscriptaddress,prl,preprintnumbers,letterpaper,floatfix,showpacs]{revtex4-1}
\usepackage{gensymb} 
\usepackage{amsmath, amssymb}
\usepackage{graphicx}
\usepackage{times}
\usepackage{upgreek}
\usepackage{dcolumn}
\usepackage{hyperref}
\usepackage{color}
\hypersetup{colorlinks = true,linkcolor = blue,anchorcolor =red,citecolor = blue,filecolor = red,urlcolor = red,
	pdfauthor=author}

\begin{document}


\title{Muon Spin Relaxation Study of frustrated Tm$_3$Sb$_3$Mg$_2$O$_{14}$ with kagom\'{e} lattice}
\author{Yanxing Yang}
\author{Kaiwen Chen}
\author{Zhaofeng Ding}
\affiliation{State Key Laboratory of Surface Physics, Department of Physics, Fudan University, Shanghai 200438, China}
\author{Adrian~D. Hillier}
\affiliation{ISIS Facility, STFC Rutherford Appleton Laboratory, Chilton, Didcot, Oxfordshire, OX110QX, United Kingdom}

\author{Lei Shu}
\email{leishu@fudan.edu.cn}
\affiliation{State Key Laboratory of Surface Physics, Department of Physics, Fudan University, Shanghai 200438, China}
\affiliation{Collaborative Innovation Center of Advanced Microstructures, Nanjing 210093, China}
\affiliation{Shanghai Research Center for Quantum Sciences, Shanghai 201315, China}

\date{\today}

\begin{abstract}
The structure and magnetic properties of rare-earth ions Tm$^{3+}$ kagom\'{e} lattice Tm$_3$Sb$_3$Mg$_2$O$_{14}$ are studied by X-ray diffraction, magnetic susceptibility and muon spin relaxation ($\mu$SR) experiments. The existence of a small amount of Tm/Mg site-mixing disorder is revealed. DC magnetic susceptibility measurement shows that Tm$^{3+}$ magnetic moments are antiferromagnetically correlated with a negative Curie-Weiss temperature of -26.3~K. Neither long-range magnetic order nor spin-glass transition is observed by DC and AC magnetic susceptibility, and confirmed by $\mu$SR experiment down to 0.1~K. However, the emergence of short-range magnetic order is indicated by the zero-field $\mu$SR experiments, and the absence of spin dynamics at low temperatures is evidenced by the longitudinal-field $\mu$SR technique. Compared with the results of Tm$_3$Sb$_3$Zn$_2$O$_{14}$, another Tm-based kagom\'{e} lattice with much more site-mixing disorder, the gapless spin liquid like behaviors in Tm$_3$Sb$_3$Zn$_2$O$_{14}$ can be induced by disorder effect. Samples with perfect geometrical frustration are in urgent demand to establish whether QSL exits in this kind of materials with rare-earth kagom\'{e} lattice.
\end{abstract}

\maketitle

\section{INTRODUCTION} \label{sec:intro}

Quantum spin liquid (QSL) is one of the most exotic states in quantum materials~\cite{Savary2016,Keimer2017,Zhou2017}. Due to the effect of quantum fluctuations, the highly entangled spins in a QSL state keep dynamic disordered state even at zero temperature. In the 1980's, Anderson proposed that a QSL can be converted into a superconductor by doping~\cite{Anderson1987}. This has attracted many efforts devoted in this area. Nowadays, the potential application in quantum computing and quantum information brings more attentions to QSL~\cite{Kitaev2006}. However, to realize a QSL state in real materials is very challenging since the defects, such as site-mixing, magnetic or non-magnetic impurities, in materials are unavoidable~\cite{disorder_Furukawa2015,disorder_Savary2017,Kimchi2018NC,disorder_Ma2018,disorder_Parker2018}.

Materials with strong magnetic frustration provide a good platform to discover QSL state~\cite{Steven2001,Balents2010,Broholm2020}. Kagom\'{e} lattice is constructed by corner-shared triangles, and theoretically has a larger ground state degeneracy than the normal triangular lattice\cite{Sachdev1992,Mila1998}. The frustrated antiferromagnetic kagom\'{e} lattice is a promising choice of seeking QSL state. The herbertsmithite ZnCu$_3$(OH)$_6$Cl$_{2}$ is a representative kagom\'{e} QSL candidate, which has been extensively studied~\cite{Herb_Bert2007,Herb_Helton2007,Herb_Imai2008,Herb_Mendels2007,Herb_Shore2005}. The magnetic susceptibility measurements show a lager negative Curie-Weiss temperature of  $\sim$-300 K, indicating a strong antiferromagnetic correlation. The absence of magnetic order~\cite{Herb_Helton2007,Herb_Mendels2007}, the low-energy gapless excitations, and the spin-excitation continuum~\cite{Herb_de2009} indicate the existence of QSL state. However, the compound cannot be made with less than a few percent Zn/Cu site disorder. In recent years, a series of new kagom\'{e} lattice $RE_3$Sb$_3$Zn$_2$O$_{14}$ have been discovered~\cite{Sanders2016,Ding2018,Ma2020}, where $RE$ represents rare-earth elements constructing the kagom\'{e} net, and most members keep disordered down to at least 2~K. Among them, Tm$_3$Sb$_3$Zn$_2$O$_{14}$ has attracted more attention~\cite{Ding2018,Ma2020}. The possible gapless spin liquid state was proposed based on the observations of the absence of magnetic order down to 20~mK, a clear linear-$T$ temperature dependence of magnetic specific heat, and the plateau behavior of temperature dependence of dynamic muon spin relaxation rate at low temperatures~\cite{Ding2018}. However, a large amount of Tm/Zn site-mixing disorder was also found in this compound, therefore, the possibility of disorder induced exotic properties cannot be ruled out~\cite{Ding2018}. By replacing Zn with Mg, the site-mixing disorder in the isostructural compound Tm$_3$Sb$_3$Mg$_2$O$_{14}$ is much reduced, the residual linear term in specific heat and the intensity of the gapless spin excitations in the inelastic neutron scattering spectra were also found to be reduced in Tm$_3$Sb$_3$Mg$_2$O$_{14}$ ~\cite{Ma2020}. This indicates that the observed properties in Tm$_3$Sb$_3$Zn$_2$O$_{14}$ are highly correlated to Tm/Zn site-mixing disorder.

To investigate whether the absence of magnetic order and persistent spin dynamics in Tm$_3$Sb$_3$Zn$_2$O$_{14}$ revealed by $\mu$SR  exist in Tm$_3$Sb$_3$Mg$_2$O$_{14}$, we performed $\mu$SR experiments on polycrystalline sample of Tm$_3$Sb$_3$Mg$_2$O$_{14}$. The XRD result shows that about 2$\%$ Tm sites are randomly occupied by Mg atoms. Magnetic susceptibility measurements exclude the existence of both long-range magnetic order and spin-glass transition. These results are consistent with a previously reported work~\cite{Ma2020}. The presence of short-range magnetic order is captured by the initial asymmetry loss behavior from the zero field (ZF)-$\mu$SR spectra. Besides, the absence of persistent spin dynamics is evidenced by the longitudinal field (LF)-$\mu$SR decoupling experiment. The $\mu$SR results are different from that in Tm$_3$Sb$_3$Zn$_2$O$_{14}$~\cite{Ding2018}, giving evidence that the disorder effect plays a crucial role in controlling the spin state in these materials.

\section{EXPERIMENTS} \label{sec:exp}

Polycrystalline Tm$_3$Sb$_3$Mg$_2$O$_{14}$ was synthesized by solid state reaction method. Stoichiometric amounts of Tm$_2$O$_3$, Sb$_2$O$_3$, and MgO were thoroughly mixed and grounded before being heated at 1350~\celsius~(or 1100~\celsius) for 7 days with several regrinding and reheating. The powder XRD data were obtained by using a Bruker D8 advanced X-ray diffraction spectrometer ($\lambda$ = 1.5418 \AA) at room temperature. The XRD Rietveld refinement was conducted by using GSAS~\cite{GSAS} program and EXPGUI~\cite{EXPGUI}. DC magnetic susceptibility measurements down to 1.8~K were carried out on a SQUID magnetometer (Quantum Design). AC magnetic susceptibility from 0.05~K to 4~K was measured by using the Physical Properties Measurement System (Quantum Design) equipped with a dilution refrigerator. $\mu$SR experiments were carried out on MuSR spectrometer at ISIS Neutron and Muon Facility, Rutherford Appleton Laboratory Chilton, UK. Muon is an extremely sensitive local magnetic probe with a resolution of 0.1~G. In a single measurement, millions of 100$\%$ spin-polarized muons are implanted into the sample and precess in the local fields at their stopping sites. After a short period of time (life time of muon is $\sim$2.2~$\mu$s), muons decay and emit a positron preferentially along the muon spin direction. By detecting the counting rate $N(t)$ in two opposite directions, we can obtain the $\mu$SR asymmetry time spectrum $A(t) = [N_F(t) - \alpha N_B(t)]/[N_F(t) + \alpha N_B(t)]$, where $N_F(t)$ and $N_B(t)$ are the counting rates of the forward and backward positron detectors with reference to the initial spin directions at time $t$. The ensemble muon spin polarization function is $P(t) = A(t)/A_0$, where $A_0$ is the initial asymmetry $A(t=0)$~\cite{Hillier2022}. The $\mu$SR experiments of Tm$_3$Sb$_3$Mg$_2$O$_{14}$ were carried out down to 0.1~K in zero field and in longitudinal field up to 30~mT.

\section{RESULTS} \label{sec:results}

\subsection{\boldmath A. Crystal structure} \label{sec:struc}
\begin{figure}[t]
 \begin{center}
 \includegraphics[width=\columnwidth]{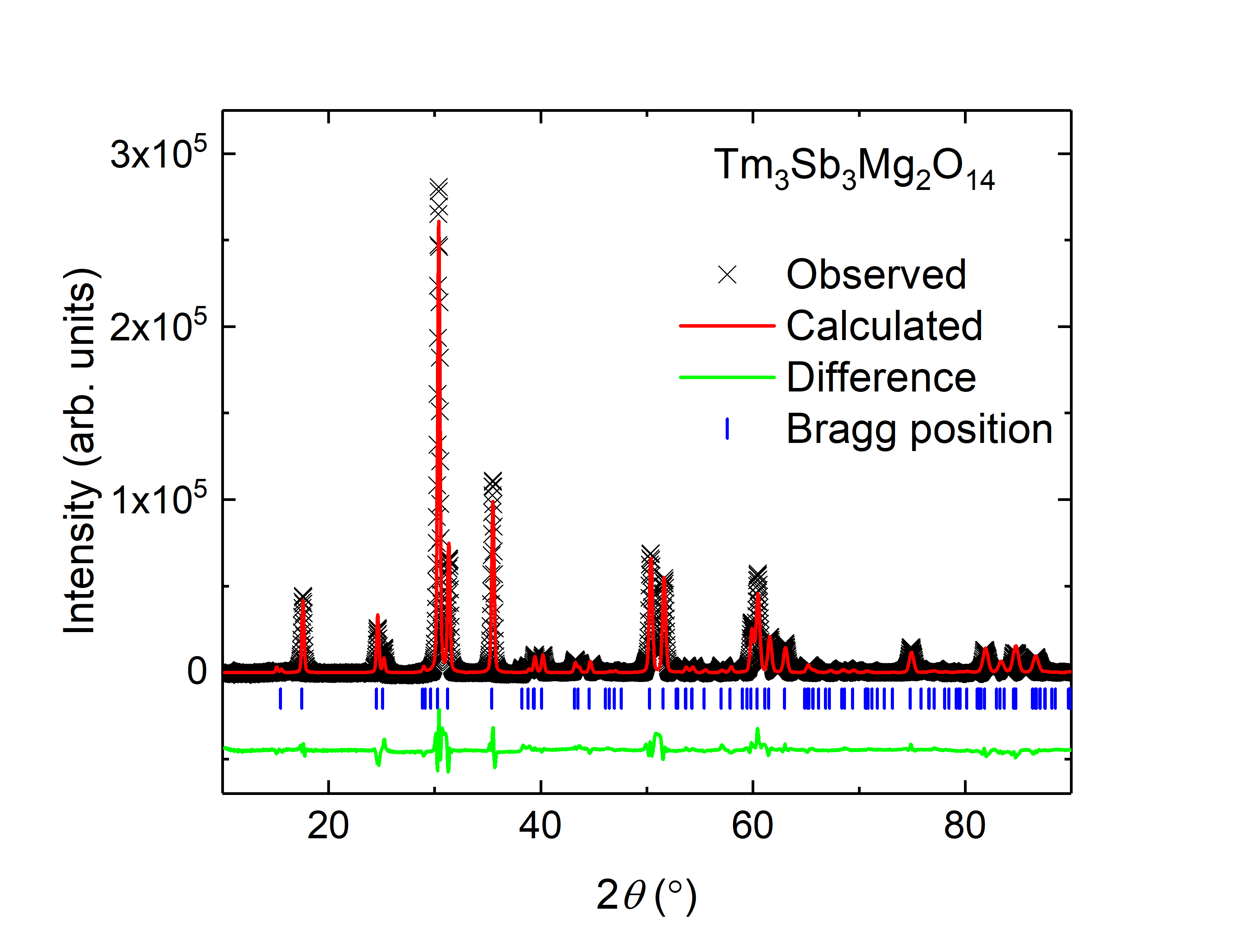}
 \caption{ Powdered XRD pattern of Tm$_3$Sb$_3$Mg$_2$O$_{14}$ measured at room temperature.}
 \label{fig:XRD}
 \end{center}
\end{figure}

In Tm$_3$Sb$_3$Mg$_2$O$_{14}$, there are two types of kagom\'{e} layers consisting of magnetic Tm$^{3+}$ ions or nonmagnetic Sb$^{5+}$ ions. These two layers alternate along $c$-axis and obey an ABC stacking arrangement. The bivalent Mg$^{2+}$ ions reside in the center of hexagons in the kagom\'{e} lattice. In the $RE_3$Sb$_3$Zn$_2$O$_{14}$ ($RE$ = rare-earth elements) family, $RE$/Zn site mixing is a common problem. In Tm$_3$Sb$_3$Zn$_2$O$_{14}$, for example, the small-angle diffraction peaks in XRD pattern almost disappear, indicating that the crystal symmetry is lowered. Further XRD refinement shows that Tm (9e) and Zn (3a) atoms in Tm$_3$Sb$_3$Zn$_2$O$_{14}$ are almost randomly mixed~\cite{Ding2018,Ma2020}.  For the powder XRD pattern of Tm$_3$Sb$_3$Mg$_2$O$_{14}$, the small-angle diffraction peaks get recovered compared with that of Tm$_3$Sb$_3$Zn$_2$O$_{14}$, as shown in Fig.~\ref{fig:XRD}. This indicates that the site-mixing in Tm$_3$Sb$_3$Mg$_2$O$_{14}$ is limited and the lattice asymmetry is unchanged. Therefore, the site-mixing (Tm/Mg) ratio is greatly reduced by replacing Zn with Mg. Table~\ref{table-A} shows the detailed results of the best refinement. The relatively large value of $\chi^2$ and $R_{wp}$  is due to the high statistics XRD data~\cite{Toby06} in order to obtain the accurate value of site-mixing fraction. In Tm$_3$Sb$_3$Mg$_2$O$_{14}$, only about 2$\%$ Tm atoms leave their original positions and interchange with Mg1, and it is a good reference for studying the disorder effect in Tm$_3$Sb$_3$Zn$_2$O$_{14}$.

\begin{table} [t]
	\begin{center}
		\caption{Rietveld refinement results for Tm$_3$Sb$_3$Mg$_2$O$_{14}$.
			$R_{wp}=14.37\%, R_{p}=9.08\%, \chi^2=13.0$; 
			the lattice constants $a=b=7.259$~\AA, $c= 17.189$~\AA; 
			$\alpha=\beta=90^{\circ}$, $\gamma=120^{\circ}$; Space Group: $R-3m$.} \label{table-A}
		\begin{tabular}{ccccccc}
			\hline
			\hline
			& Wyckoff \\
			Atom & positions & $x$ & $y$ & $z$ & B/\AA$^2$ & Occ.\\
			\hline
			Tm & 9e & 0.5 & 0 & 0& 0.004 & 0.98 \\
			Mg(disorder) & 9e & 0.5 & 0 & 0& 0.025 & 0.02 \\					
			Mg1 & 3a & 0 & 0 & 0 & 0.016 & 0.94 \\
			Tm(disorder) & 3a & 0 & 0 & 0 & 0.025 & 0.06 \\
			Mg2 & 3b & 0 & 0 & 0.5 & 0 & 1 \\
			Sb & 9d& 0.5 & 0 & 0.5 & 0.008 & 1 \\
			O1 & 6c & 0 & 0 & 0.0397(2) & 0.074 & 1 \\
			O2 & 18h & 0.5123(1) &-0.5123(1) & 0.1348(7) & 0.011 & 1 \\
			O3 & 18h & 0.1564(2) &-0.1564(2) &-0.0535(2) & 0.026 & 1 \\				
			\hline
			\hline
		\end{tabular}
	\end{center}

\end{table}

\subsection{\boldmath B. Magnetic susceptibility} \label{sec:sus}
\begin{figure}[t]
	\begin{center}
		\includegraphics[width=7cm]{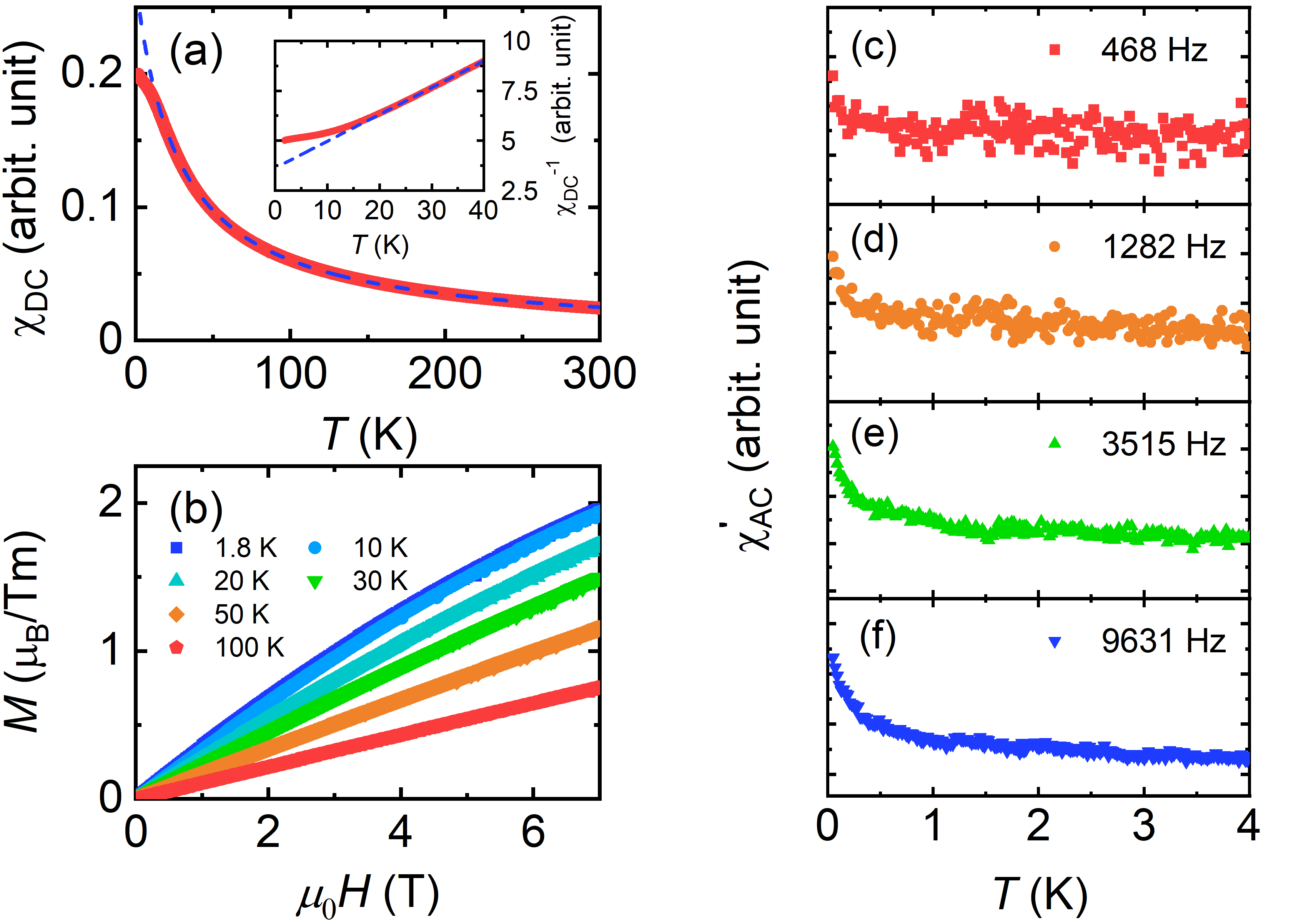}
			\caption{\textbf{(a)} Temperature dependence of magnetic susceptibility $\chi_{\rm DC}$ of polycrystalline Tm$_3$Sb$_3$Mg$_2$O$_{14}$ under an applied magnetic field of $\mu_0H=0.1$~T. The inset shows the inverse of magnetic susceptibility $1/\chi_{\rm DC}$ at low temperatures. The blue dashed line is the Curie-Weiss law fitting curve. \textbf{(b)} Isothermal magnetization between temperature 1.8~K and 100~K, applied magnetic fields up to 7~T. \textbf{(c)}-\textbf{(f)} Temperature dependence of real part of AC susceptibility $\chi_{\rm{AC}}'$ with different driving frequencies in zero static field down to $T=50$~mK.}
		\label{fig:chi}
	\end{center}
\end{figure}

The DC magnetic susceptibility $\chi_{\rm {DC}}$ of Tm$_3$Sb$_3$Mg$_2$O$_{14}$ was measured under the magnetic field of $\mu_0H=0.1$~T, as shown in Fig.~\ref{fig:chi}(a). There is no sign of long-range magnetic order or spin glass behavior down to 1.8~K, which is consistent with previous work~\cite{Ma2020}. The inverse susceptibility $1/\chi_{\rm{DC}}$ deviates from the straight line slightly when cooling across 20~K. As a result, Curie-Weiss fitting $\chi_{\rm{DC}} = C/(T-T_{\rm{CW}}) + \chi_0$ was implemented in the temperature range between 20~K and 300~K. The Curie-Weiss temperature $T_{\rm{CW}}$ is -26.3~K, indicating the antiferromagnetic correlation between the non-Kramers Tm$^{3+}$ ions. The calculated effective moment $\mu_{\rm{eff}}$
 is 7.53 $\mu_{\rm{B}}$, close to the theoretical value of 7.57 $\mu_{\rm B}$ for  the electron configuration $4f^{12}$ of Tm$^{3+}$~\cite{Ashcroft76}. Fig.~\ref{fig:chi}(b) shows the magnetization curves at different temperatures. In a wide temperature range, the saturation tendency is not apparent up to 7~T, indicating the existence of strong magnetic frustrations, or the interaction among Tm$^{3+}$ ions in this material are relatively strong. AC magnetic susceptibility $\chi_{\rm{AC}}$ of Tm$_3$Sb$_3$Mg$_2$O$_{14}$ was measured in zero static field with different driven frequencies. As shown in the right panels of Fig.~\ref{fig:chi}, no peaks or frequency dependency is observed, indicating that spin glass transition is not found down to $T=50$~mK.

\subsection{\boldmath C. $\mu$SR} \label{sec: uSR}

\begin{figure}[t]
	\begin{center}
		\includegraphics[width=7cm]{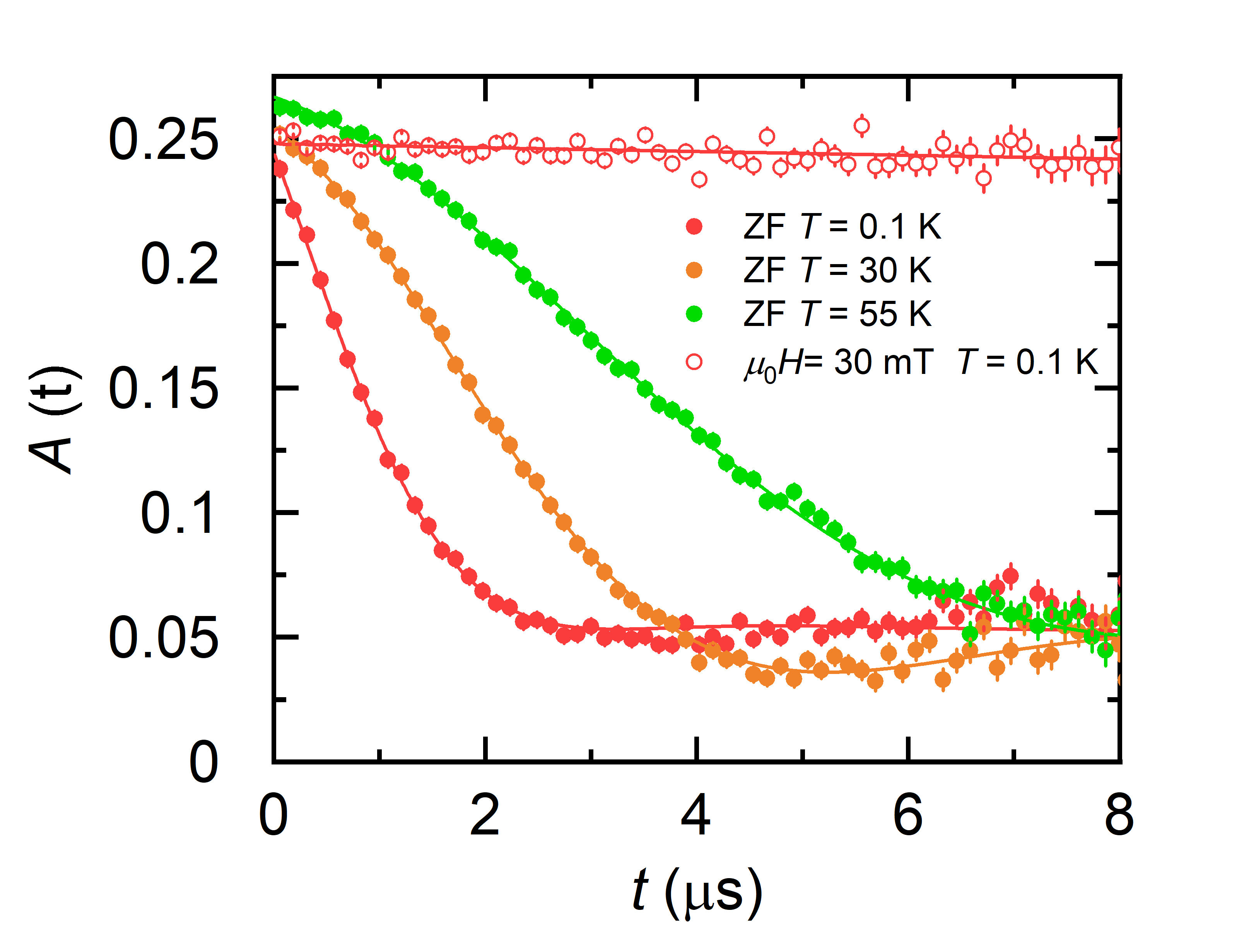}
		\caption{$\mu$SR spectra of Tm$_3$Sb$_3$Mg$_2$O$_{14}$ measured at different temperatures and fields. Red to Green points: ZF-$\mu$SR spectra down to 0.1~K. The solid lines stand for the fitting with Eq.~(\ref{eq:1}). Red open circles: LF-$\mu$SR spectrum measured under 30~mT. The solid line stands for the fitting with Eq.~(\ref{eq:3}).}
		\label{fig:musr}
	\end{center}
\end{figure}

Since muons are extremely sensitive to local magnetic moments, ZF-$\mu$SR is a powerful technique to detect long-range magnetic order or spin freezing~\cite{Zhu2020,Hillier2022}.  Representative ZF-$\mu$SR spectra measured at different temperatures are shown in Fig.~\ref{fig:musr}. No oscillations were observed, suggesting the absence of long-range magnetic order down to 0.1~K. ZF-$\mu$SR data after subtracting the background signal can be well fitted with a zero field Kubo-Toyabe (ZF-KT) function times an exponential term:
\begin{eqnarray}
	\label{eq:1}
	A_{\rm {ZF}}^{\rm{KT}}(t)=A_0(T)G_{\rm{ZF}}^{\rm{KT}}(\sigma,t)e^{-\lambda t},
\end{eqnarray}
The ZF-KT term:
\begin{eqnarray}
	\label{eq:2}
	G_{\rm{ZF}}^{\rm{KT}}(\sigma,t)=\frac{1}{3}+\frac{2}{3}(1-\sigma^2t^2)\cdot\exp(-\frac{1}{2}\sigma^2t^2)
\end{eqnarray}
describes the muon spin relaxation due to randomly oriented static local fields of Gaussian distribution with distribution width $\delta B_{\rm{G}}=\sigma/\gamma_\upmu$, 
where $\sigma$ is relaxation rate and $\gamma_\upmu=2\pi\times135.53$~MHz/T is the muon gyromagnetic ratio. The exponential term stands for the contribution of dynamic relaxation or static local fields with lorentzian distribution.  $\emph{A}$$_0$($\emph{T}$) is the temperature dependent initial symmetry. 

\begin{figure}[h]
	\begin{center}
		\includegraphics[width=\columnwidth]{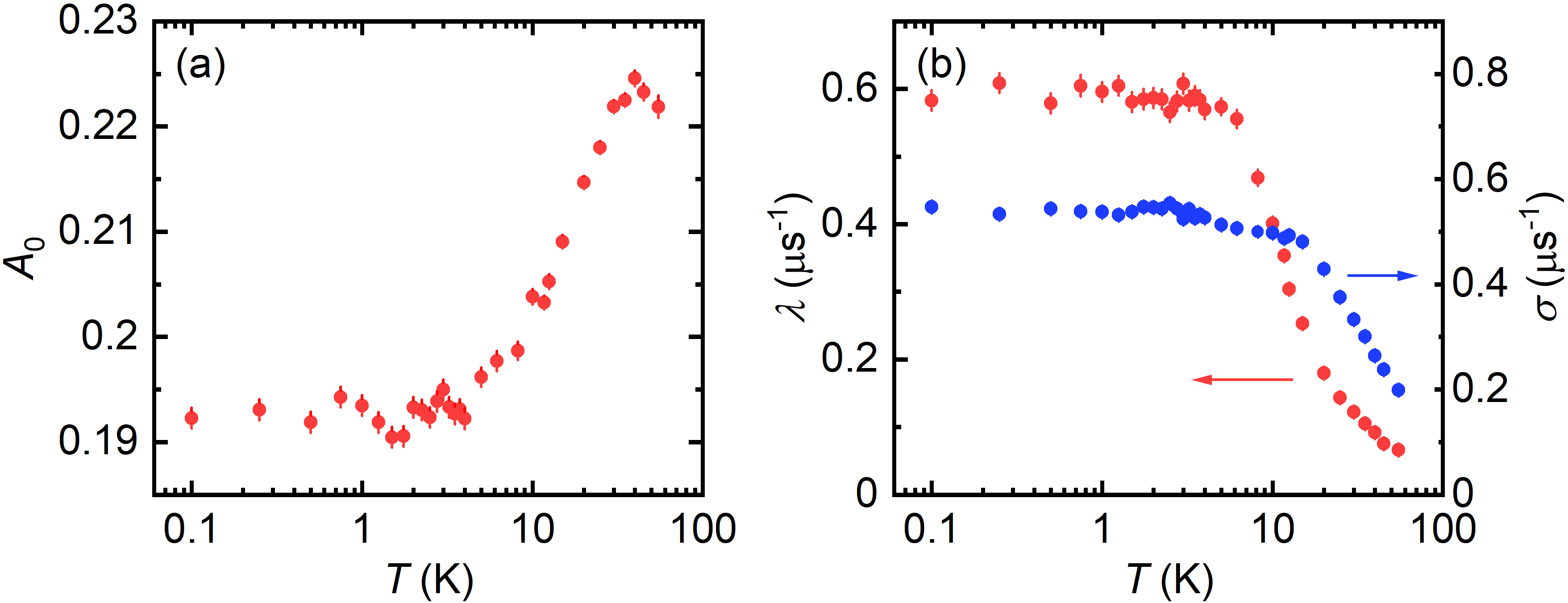}
		\caption{\textbf{(a)} Temperature dependence of ZF-$\mu$SR initial asymmetry $\emph{A}$$_0$. \textbf{(b)} Temperature dependence of muon spin relaxation rates $\lambda$ and $\sigma$.}
		\label{fig:musr-para}
	\end{center}
\end{figure}

The temperature dependence of fitting parameters $A_0$, $\sigma$, and $\lambda$ are shown in Fig.~\ref{fig:musr-para}. The decrease of $A_{\rm 0}$ with decreasing temperature is due to the onset of short-range magnetic order. There is a strong local field in a fraction of the sample volume, so that muons in this volume are rapidly depolarized and do not contribute to the signal~\cite{de1997muon}. This “lost” volume fraction $1-A_{\rm 0}(T) / A_{\rm 0}(50~\rm K)$, where $A_{\rm 0}(50~\rm K)$ is the initial asymmetry at $T = 50$~K, increases with decreasing temperature to about 15\% at low temperatures in Tm$_3$Sb$_3$Mg$_2$O$_{14}$. The temperature where $A_{\rm 0}$ begins to decrease is nearly consistent with the temperature where the deviation of Curie-Weiss law below 20 K (inset of Fig.~{\ref{fig:chi}}(a)).  

Both relaxation rates $\sigma$ and $\lambda$ gradually increase with decreasing temperature and saturate at low temperatures. The static relaxation rate $\sigma$ usually stems from nuclear dipolar moment. $\sigma$ gradually increases as temperature is cooled down and finally saturates below 3~K. A similar temperature dependence of $\sigma$ was also found in the sandglass magnet Tm$_3$SbO$_7$~\cite{Yang2022}, where $\sigma$ is proportional to $\chi_{\rm {DC}}$ in a wide temperature range, and the hyperfine-enhanced Tm nuclear moments were proposed. For Tm$_3$Sb$_3$Mg$_2$O$_{14}$, $\sigma$ does not scale with $\chi_{\rm DC}$ well below 20~K due to the emergence of short-range order.

When applying a longitudinal field $B=\omega_0/\gamma_\upmu$, $G_{\rm{ZF}}^{\rm{KT}}(\sigma,t)$ in Eq.~(\ref{eq:1}) changes into the LF-KT form:
\begin{multline}
	\label{eq:3}
	G_{\rm {LF}}^{\rm {KT}}(\sigma,t)=1-\frac{2\sigma^2}{\omega_0^2}[1-e^{-\frac{1}{2}\sigma^2t^2}\rm {cos}(\omega_0t)]\\
	+\frac{2\sigma^4}{\omega_0^3}\int_{0}^{t}e^{-\frac{1}{2}\sigma^2\tau^2}\rm {sin}(\omega_0\tau)d\tau,
\end{multline}
where $\sigma$ is the same as in ZF-KT function. Typically, for applied LF 10 times larger than the static local fields at muon sites, the resultant static field is nearly parallel to muon spin polarization, so there is little static muon relaxation. Then muon spin polarization function is said to be ``decoupled'' from the static local fields. However, if the internal magnetic field is dynamic, the relaxation is not easily suppressed by the applied longitudinal filed. The LF-$\mu$SR spectrum of 30~mT is also shown in Fig.~\ref{fig:musr}. The relaxation function is fully decoupled under the relatively low field, indicating the static origin of the exponential term $e ^{-\lambda t}$ in Eq.~(\ref{eq:1}). We notice that the rapid increase of $\lambda$ below 20~K is accompanied by the decrease of initial asymmetry $A_{\rm 0}$, which indicates the appearance of short-range magnetic order. And both $\lambda$ and $A_{\rm 0}$ become temperature independent at the same temperature. It is reasonable to argue that part of the Gaussian field distribution of nuclear dipole moment at muon site is changed to Lorentzian distribution due to the appearance of short-range magnetic order.

\section{DISCUSSIONS} \label{sec:discussion}
The XRD refinement shows that there are only about 2\% Tm/Mg site-mixing in Tm$_3$Sb$_3$Mg$_2$O$_{14}$. This is very different from that of Tm$_3$Sb$_3$Zn$_2$O$_{14}$, in which 17.5\% disorder was observed~\cite{Ding2018}. However, for both compounds, the magnetic susceptibility measurements do not show any signatures of magnetic phase transition or spin freezing. From the specific heat measurements, a linear residual term $\gamma$ was found in Tm$_3$Sb$_3$Mg$_2$O$_{14}$~\cite{Ma2020}, whose magnitude is about 3 times smaller than that of Tm$_3$Sb$_3$Zn$_2$O$_{14}$~\cite{Ma2020,Ding2018}. A similar behavior was also found in the inelastic neutron scattering measurement, i.e., the intensity of the magnetic excitations in Tm$_3$Sb$_3$Mg$_2$O$_{14}$ is reduced compared to that of Tm$_3$Sb$_3$Zn$_2$O$_{14}$~\cite{Ma2020}. While the absence of long-range magnetic order and persistent spin dynamics were indicated in Tm$_3$Sb$_3$Zn$_2$O$_{14}$ by $\mu$SR measurements~\cite{Ding2018},  short-range magnetic order is revealed based on the observation of the loss of initial asymmetries, and the LF decoupling experiment excludes the existence of persistent spin dynamics in Tm$_3$Sb$_3$Mg$_2$O$_{14}$. Therefore, for both Tm$_3$Sb$_3$Zn$_2$O$_{14}$ with large amount of site-mixing, and Tm$_3$Sb$_3$Mg$_2$O$_{14}$ with only 2\% site-mixing samples, QSL ground state does not exit. Samples with perfect geometrical frustration are in urgent demand to establish whether QSL exits in this kind of materials with rare-earth kagom\'{e} lattice.

For Tm$_3$Sb$_3$Mg$_2$O$_{14}$, the 13-fold degeneracy of the total moment of Tm$^{3+}$ ($J = 6$) is reduced by crystal electric field (CEF). The ground state is a non-Kramers doublet, which can be regarded as an effective spin-$1/2$ moment~\cite{Ding2018,Ma2020}. Magnetization measurements indicates that there are strong exchange interactions between the neighboring spins. Tm$_3$Sb$_3$Mg$_2$O$_{14}$ with much less disorder seems to be a good QSL candidate. However, short-range magnetic order was observed by ZF-$\mu$SR experiment. This can be due to the 2\% Tm/Mg site-mixing disorder. A similar phenomenon has also been found in the Yb-triangle lattice  NaYbSe$_2$ with about 4-5\% Na vacancies~\cite{Dai2021}, in which about 23\% quasi-static spins were found by $\mu$SR and confirmed by NMR experiment~\cite{Zhu2021}. 

The specific heat and inelastic neutron scattering results give evidence that the low-energy density of state observed in Tm$_3$Sb$_3$Mg$_2$O$_{14}$ and Tm$_3$Sb$_3$Zn$_2$O$_{14}$ is related to the disorder. A theoretical model that disorder can induce a low-energy density of state in the Kitave QSL system~\cite{Knolle2019,Kao2021} was recently proposed. Whether disorder can also induce the low-energy density of state in the geometry frustrated magnetic system may also worth further study. 

\section{CONCLUSIONS} \label{sec:conc}
XRD, magnetic susceptibility, and $\mu$SR experiments have been performed on the Tm$^{3+}$ kagom\'{e} lattice Tm$_3$Sb$_3$Mg$_2$O$_{14}$. The existence of about 2\% Tm/Mg site-mixing disorder is revealed by the XRD Rietveld refinement. Neither long-range magnetic order nor spin-glass transition is observed by DC and AC magnetic susceptibility measurements, and confirmed by $\mu$SR experiments down to 0.1~K. However, the emergence of short-range magnetic order is discovered by ZF-$\mu$SR experiments, and the absence of spin dynamics at low temperatures is evidenced by LF-$\mu$SR technique. All of these observations suggest that QSL ground state does not exist in our Tm$_3$Sb$_3$Mg$_2$O$_{14}$ compound. Compared with the results of Tm$_3$Sb$_3$Zn$_2$O$_{14}$, another Tm-based kagom\'{e} lattice with much more site-mixing disorder, the gapless spin liquid like behaviors in Tm$_3$Sb$_3$Zn$_2$O$_{14}$ can be induced by disorder effect. Realization of perfect frustrated structure in real material is very important on the way to pursue QSL ground state.

 {\textit{Acknowledgments --} We are grateful to the ISIS cryogenics Group for their valuable help during the $\mu$SR experiments (10.5286/ISIS.E.RB1820271). This research was funded by the National Natural Science Foundations of China, No. 12034004 and 12174065, and the Shanghai Municipal Science and Technology (Major Project Grant No. 2019SHZDZX01 and No. 20ZR1405300). }

%


\end{document}